\documentstyle[12pt]{article}

\renewcommand{\theequation}{\arabic{equation}}
\def\sqr#1#2#3#4{{\vcenter{\vskip -#3 pt\hbox{\kern #4 pt\vbox{\hrule height
.#2 pt\hbox{\vrule width .#2 pt height #1 pt\kern #1 pt\vrule width .#2 pt}
\hrule height .#2 pt}\kern #4 pt}}}}

\begin{document}

\begin{titlepage}
\begin{flushright}
HEP-TH/9410231\\
IPM-94-67 \\
SUTDP/94/73/10 \\
Oct. 1994
\end{flushright}

\begin{center}

{\bf CHIRAL PERTURBATION THEORY IN THE FRAMEWORK OF NON-COMMUTATIVE GEOMETRY
}
\vspace{1.0cm}

{\bf  Farhad  Ardalan
\footnote{ardalan@irearn.bitnet}}
and {\bf  Kamran Kaviani
\footnote{kaviani@irearn.bitnet}}\\

{\it  Institute for Studies in Theoretical Physics and Mathematics,\\
P.O.Box 19395-5746 Tehran-Iran.}\\
and\\
{\it
    Dept. of Physics,
    Sharif University of Technology,\\
    P.O. Box: 11365-9161, Tehran, Iran}\\

\end{center}


\abstract{
We consider the non-commutative generalization of the chiral perturbation
theory. The resultant coupling constants are severely restricted by the model
and in good agreement with the data. When applied to the Skyrme model,
our scheme reproduces the non-Skyrme term with the right coefficient.
We comment on a similar treatment of the linear $\sigma $-model.
}
\vfill
\end{titlepage}


\section{Introduction}

Discovery of Non-Commutative Geometry (NCG) by Alain Connes \cite{C1} and its
subsequent application to problems in particle physics has attracted
considerable attention in recent years \cite{C2,C3,K1,K2,BG,SZ}. In its
simplest form,
the generalized gauge fields of the non-commutative version of the electroweak
sector of the standard model naturally included the Higgs field whose origin in
the standard model is obscure, to say the least. The mass scale of the Higgs
sector is directly connected to the geometrical separation parameter of the
NCG gauge theory. Other parameters of the standard model such as mass ratios
are also determined in this construction \cite{C1}. Subsequently, the
non-commutative geometric construction
has been applied to the extensions of the standard model \cite{K1,K2},
to grand unified field
theories \cite{Ch1,Ch2}, and even to gravitational interactions \cite{Ch3}.
In the former theories,
 the non-commutative component gives rise to spontaneous symmetry breaking and
its Higgs field, while in the latter case a dilaton field emerges from the
generalized gravity construction. There is one area however, in which the
application of non-commutative framework has not proven
straightforward, and
that is in QCD. A simple minded procedure will result in an unwanted color
symmetry break down \cite{K2}. It is entirely possible that more intricate use
of non-commutative geometry
ideas may tame QCD into its framework, just as a truly non-commutative
quantum theory of gravity has eluded us yet, an area in which the primary
expectation of non-commutative geometric phenomena lie.

But then there is one region of QCD which lends itself to a similar treatment
as in the
original application of non-commutative geometry to electroweak sector of
standard model. The spontaneous breakdown
of the chiral flavour symmetry of QCD in low energies invites a non-commutative
exploration. We will take up the application of non-commutative geometric
structure to the chiral effective Lagrangian of low energy QCD in this paper.

Chiral perturbation theory (ChPT) as an effective field theory for low
energy QCD has
been applied successfully to a wide range of problems in hadron physics
\cite{DD}.
In the simplest form ChPT involves the pseudoscalar mesons as the basic
fields . Interaction of these fields is described by an effective Lagrangian
which is ordered with respect to the number of derivatives and mass of the
meson
field. To lowest order i.e. second order in momenta, it simply is the
non-linear
$\sigma $-model which was applied successfully to low energy hadron physics in
the early days of current algebra \cite{LNG}. To fourth order in derivatives,
it involves a rather large number of terms, which have enabled a detailed and
quantitative analysis of such quantities as form factors and
scattering amplitudes
for various meson processes \cite{GL1,GL2}.

A simple elegant chiral perturbative
Lagrangian
which involves some quartic terms is the Skyrme model, which has been studied
extensively. It describes not only the mesons as the basic fields,
 but also baryons as the soliton solutions, of the theory.

In its application to
hadron physics, it was realized \cite{DG} that an extra term, the so called
non-Skyrme
or symmetric quartic term  has to be added, with an adjustable coefficient
to describe $\pi \pi $ scattering correctly.

 In this paper we will generalize the chiral lagrangian to a two sheeted
non-commutative geometry. To second order in momentum, we get the non
linear $\sigma $-model with a mass term appearing as a result of the
non-commutativity of our geometry. To fourth order we obtain the usual
chiral perturbation theory lagrangian, with the coupling constants severely
restricted. Comparison with the data shows fairly good agreement. In the
special region of Skyrme model validity, we obtain both the Skyrme lagrangian
and the non-Skyrme term, again with its coefficient determined by the theory
and in good agreement with the data.

In section 2, we will review the non-commutative geometry construction and
remind the reader of the original application to standard model, thereby
setting up our formulation and notation.
In section 3, we will apply the non-commutative procedure first to the case of
the non-linear $\sigma$-model, to the conventional ChPT, and then to
the case of the standard Skyrme model.
In the appendix we will study the linear $\sigma$-model in the
framework of
non-commutative geometry. \footnote{ After this work was completed we became
aware of a preprint \cite{G} which also studied the linear sigma model in
the context of non-commutative geometry, but with a different method.}

\section{Review of Non-Commutative Geometry  }

In this section we remind some of the relevant features of the non-commutative
geometry and set our notation. For detail and further analysis the reader
is refered to the references \cite{C1,C2,C3}.
The basic objects of such
a geometry is a possibly non-commutative algebra ${\cal A}$ which is
the generalization of the algebra of functions
on a manifold, and a Dirac k-cycle which is a triplet (${\cal H},D,\Gamma$),
where ${\cal H} $ is a Hilbert space, $D$ is the Dirac operator and $\Gamma $
is a ${ Z\kern - .5em Z}_2 $ grading operator. The algebra $ {\cal A} $ is an
associative algebra with unit
$1$ and an involution $*$. A (matrix) representation of
${\cal A}$ over
the Hilbert space ${\cal H}$ is a homomorphism from ${\cal A}$ into the linear
operators on ${\cal H}$ which is faithful. Given an involutive algebra ${\cal
A}$ which corresponds to a geometrical space, we would like to construct a
differential algebra corresponding to a differential geometry. In order to
construct this differential algebra we need
to define a linear operator $d$, as the exterior derivative, satisfying $d
^2\,=\,0$ and the Leibniz rule. Using $ d $, we can construct
the p-forms as:
\begin{equation}
\sum _i \,a_0^i\, da_1^i\, ... \,  da_p^i \,\,\,\,\,\,\,\,\,;
a_0^i,a_1^i,...,a_p^i \in {\cal A} .
\end{equation}
One defines the representation  of an element $ da$ in
${\cal H}$ as,
\begin{equation}
 da=[D,a]_g=D a - \Gamma a \Gamma D .
\end{equation}
The grading operator $\Gamma $ satisfies the following properties.
\begin{eqnarray}
\Gamma ^2 &=& 1 \nonumber \\
\Gamma \omega  &=& (-)^n \omega \Gamma ,
\end{eqnarray}
where $\omega $ is an n-form in differential algebra. In this formalism
the elements of the algebra ${\cal A} $ are taken as the 0-forms. The Dirac
operator is an unbounded self-adjoint operator in ${\cal H}$, such that
\begin{equation}
D \, \Gamma \,=\, - \, \Gamma \, D .
\end{equation}
Other differential geometric quantities such as connection and curvature
may be similarly defined \cite{C1,C2}.

Before applying the above formalism to our physical problem, let us carry out
some of calculations which will be needed in the following sections.
First we show how the differential algebra
reproduces the ordinary differential forms on a flat, compact and Euclidian
manifold ${\cal M}$. For such a manifold we should take ${\cal A} $ to be the
algebra of complex valued functions on ${\cal M}$ ,
\begin{equation}
{\cal A} := {\cal C}^\infty ({\cal M}),
\end{equation}
then $D$ is the ordinary Dirac operator
\begin{equation}
D\,=\, \partial \kern -0.5em /  \,=  \gamma ^\mu \partial _\mu
\end{equation}
and $\gamma _5 $ is the grading operator $\Gamma $.
According to (2),
\begin{equation}
dg\,\,=\,\,[\,D\,,\,g\,], \,\,\,\,\,\,\forall g \in {\cal A}
\end{equation}
which becomes,
\begin{equation}
dg=[\,\partial \kern -0.5em /  \,,\,g\,]\,\,=\,\,\gamma ^\mu\,\,\partial _\mu
\,g
\equiv \gamma ({\bf d}g)\,\,,
\end{equation}
where ${\bf d}g$ is the ordinary one-form, ${\bf d}g= \partial _\mu g {\bf
d}x^\mu $.

To describe a fiber bundle over the manifold ${\cal M }$, we take
\begin{equation}
{\cal A}={\cal C}^\infty({\cal M}) \otimes M_N({I \kern -.6em C}),
\end{equation}
where $M_N({I \kern -.6em C})$ is the algebra of $N \times N $ complex
matrices.
In this case the Dirac and grading operators are $\partial \kern -0.5em /
\otimes 1$ and
$\gamma _5 \otimes 1$ respectively.
Next we construct the differential algebra of the fiber bundle over the space
consisting of two layers, each layer a Euclidian, compact manifold ${\cal M}$.
Each layer is described by the algebra of continuous
functions ${\cal C}^\infty ({\cal M})$.
The proper algebra for this geometry is:
\begin{equation}
{\cal A}={\cal C}^\infty({\cal M}) \otimes ( \,\,M_N({I \kern -.6 em C}) \oplus
M_N({I \kern -.6em C})\,\,), \end{equation}
and the Dirac operator is \cite{Ch1}

\begin{equation}
D = \left( \matrix{ \partial \kern -0.5em /  \,\, \otimes \,\, { 1\kern - .4em
1}
& \gamma
_5 \otimes M  \cr \gamma _5 \otimes M^{\dag}  &
\partial \kern -0.5em /  \,\, \otimes \,\, { 1\kern - .4em 1} }
\right)
\end{equation}

Setting the off-diagonal terms equal
to zero, will make differentiation on one layer independent
of the other layer.

In this example we take the representation
of an elements of ${\cal A}$ to be
\begin{equation}
g = \left (\matrix {V(x) & 0 \cr 0 & V^\prime (x) } \right
)\,\,\,\,,\,\,g\,\,\in {\cal A}
 \end{equation}
where $V(x)$ and $V^\prime (x) $ are $N \times N $ matrices and $x$ indicates
the coordinates on ${\cal M}$. This representation contains in
itself the information about the two layers of space. $V(x)$ represents
functions on one layer and $V^\prime (x)$ those of the
other layer.

Now we can calculate $dg$. With the same
procedure which was used for obtaining
(8), we find
\begin{eqnarray}
dg &=& [\,\,D\,,\,g \,\,]
= \left ( \matrix{\partial \kern -0.5em /  V & \gamma _5 (M\,V^\prime
\,-\,V\,M) \cr
\gamma _5 (M^{\dag }\,V\,-\,V^\prime \,M^{\dag}) & \partial \kern -0.5em /
V^\prime} \right ).
\end{eqnarray}
In the above formula not only we have $\partial \kern -0.5em /  V$ and $
\partial \kern -0.5em / V^\prime $
 which can be related to the ordinary one-forms on each
layer of space, but also there are off-diagonal terms which are
proportional to the difference of the functions corresponding to the two
seperate layers. In fact the
$N \times N $ matrix $M$ which is called the mass matrix, establishes the
connection between the two layers.

A two  sheeted space also was used by Connes \cite{C1,C2} to reproduce the
Higgs  sector of the standard model beside the usual gauge boson sector.
For this purpose, notice that the curvature two form
$\theta =\,\,d \rho \,\,+\,\,\rho ^2 $
corresponding to the connection one-form
$\rho = \sum _i \,\,a_i\,\, db_i \,\,\,,\,\;\forall \,a_i\,,\,b_i \in {\cal A}
$
with,
 $\rho = \rho ^*$,
yields the Yang-Mills action
\begin{equation}
S_{YM}= {1 \over 4} Tr_\omega (\theta ^2 \mid D \mid ^{-4} ) ,
\end{equation}
where $Tr_\omega $ is the Dixmier trace and $\mid D \mid ^2 = D D^{\dag } $.
It can be shown in this case that the $S_{YM}$ will reduce to
\begin{equation}
S_{YM}= {1 \over 4} \int dx^4 {\sqrt g} Tr(tr(\theta ^2 )),
\end{equation}
where $tr$ is taken over the Clifford algebra and $Tr$ is taken over the matrix
structure \cite{Ch1}.

The fermionic part of the action is,
\begin{equation}
S_F=\int dx^4 {\overline \Psi} (D + \rho) \Psi .
\end{equation}
and the total action $S=S_F + S_{YM} $ can easily be shown to be
invariant under the gauge transformations,
$ \Psi \to \Psi ^\prime =g \Psi$ and
$\rho \to g \rho g^* + g dg ^*$,
with $g$ an element of the unitary group ${\cal U}$ of ${\cal A}$
\begin{equation}
g \,\,\in {\cal U}({\cal A})= \{g \in {\cal A} \,\,, gg^*=g^*g =1\},
\end{equation}
and $*$ indicating involution in ${\cal A}$.

To obtain the standard model electroweak lagrangian Connes takes the algebra to
be \begin{equation}
{\cal A}=C^{\infty}({\cal M}) \otimes ({I \kern - .6em C} \oplus { H \kern -
.8em H})
 \end{equation}
where  ${I \kern - .6em C}$ and ${ H \kern - .8em H}$, denote the
algebra of complex numbers and quaternions.
Using a Dirac operator similar to (11), the connection one form is obtained
via eq.(7) \cite{Ch1}.
\begin{equation}
\rho = \left ( \matrix { A_1 & \gamma _5 H K \cr \gamma _5 H^* K^* & A_2 }
\right )
 \end{equation}
Here $A_1$ and $A_2$ are the ordinary gauge fields over each layer
of space and $H$ is the other component of the gauge field along the
discrete dimension which acts as the Higgs field in the theory.
K is a mixing
matrix related to the fermionic mass matrix. The complete action of the
electroweak sector of the standard model is then obtained from the generalized
gauge field \cite{K1}.

\section{Chiral perturbation theory and NCG}

Chiral perturbation theory is an effective field theory of mesons,
for low energy QCD.
The strategy in this
theory is to expand the effective lagrangian in powers of the momenta and
take the lowest order terms.
To obtain the effective lagrangian, one demands that the
symmetry of this effective theory be a symmetry of QCD. As a good
approximation QCD has $ SU_L(N_f) \times SU_R(N_f) $ chiral symmetry  where
$N_f$ is the number
of massles flavours.
On the other hand Lorentz invariance forces the number of
derivatives in each term of the effective lagrangian expansion to be even.
Another important property of effective Lagrangian comes from PCAC,
which at low energies prevents the Goldstone bosons
interact with one another. Putting all these together one may write
the effective lagrangian of low energy QCD as follows \cite{L1} :
\begin{equation}
{\cal L}_{eff}= {\cal L}^{(2)}+ {\cal L}^{(4)}+ {\cal L}^{(6)} + ...
\end{equation}
${\cal L}^{(2)}$
has the form of the 4 dimensional non-linear $\sigma$-model, i.e.
\begin{equation}
{\cal L}^{(2)}={F_0 ^2 \over 4} Tr ( \partial _\mu U \partial ^\mu U^{\dag})
\,\,\, ; \,\,\,L_\mu =U \partial _\mu U ^ {\dag} \, ,\, U \in SU(N_f) ,
 \end{equation}

where $U$ is related to the mesonic field by

\begin{equation}
U= e ^ {i \tau ^a \pi _a \over F_0}
\end{equation}
and $F_0 $ is the pion decay constant.
If pions are taken to be massive, then
it is not difficult to show that ${\cal L}^{(2)}$ should be modified
as bellow \cite{L1,HS}

\begin{equation}
{\cal L}^{(2)}={F_0 ^2 \over 4} Tr (L_\mu L^\mu )+ {F_0 ^2 m_\pi ^2 \over
4 } Tr (U+U ^ {\dag} -2).
\end{equation}

Higher order terms of (20) can be written in terms of $U$ or $L_\mu
$, and increase in complexity as the order increases. Gasser and Leutwyler
\cite{GL2} have
obtained the following expression for the general Lagrangian to order $ p ^{4}
$ in the case of $N_f=3$:

\begin{eqnarray}
{\cal L}^{(4)} &=& L_1 ( Tr(\partial ^\mu U \partial _\mu U^ {\dag} ) ) ^2 +
L_2 Tr (\partial _\mu U \partial _\nu U ^ {\dag}) Tr (\partial ^ \mu U \partial
^\nu U ^ {\dag})
\nonumber \\
&+&
L_3 Tr ( \partial ^\mu U \partial _\mu U^ {\dag} \partial ^\nu U \partial _\nu
U ^{\dag}) + L_4 Tr ( \partial ^\mu U \partial _\mu U ^ {\dag})
Tr (\chi ^{\dag} U + \chi U^ {\dag})
\nonumber \\
&+&
L_5 Tr (\partial ^\mu U \partial _\mu U ^ {\dag} (\chi ^{\dag} U + U ^{\dag}
\chi )) + L_6 [Tr (\chi ^ {\dag} U + \chi U ^ {\dag})]^2
\nonumber \\
&+&
L_7 (Tr (\chi ^ {\dag} U - \chi U ^ {\dag}))^2 + L_8 Tr (\chi ^ {\dag} U
\chi ^{\dag} U + \chi U^{\dag} \chi U^{\dag})
\end{eqnarray}
In the above lagrangian we have ignored the presence of vector and axial fields
which occur in the original lagrangian and only kept the symmetry breaking
terms. The field $\chi $  contains
the information about the mesonic masses, which to lowest order is the mass
matrix $\chi = m_ \pi 1 \kern - .4
em 1$. Then the lagrangian becomes,
\begin{eqnarray}
{\cal L}^{(4)} &=& L_1 ( Tr(\partial ^\mu U \partial _\mu U^ {\dag } ) ) ^2 +
L_2 Tr (\partial _\mu U \partial _\nu U ^ {\dag}) Tr (\partial ^ \mu U \partial
^\nu U ^ {\dag})
\nonumber \\
&+&
L_3 Tr ( \partial ^\mu U \partial _\mu U^ {\dag} \partial ^\nu U \partial _\nu
U ^{\dag}) + L_4 m_\pi ^ 2 Tr ( \partial ^\mu U \partial _\mu U ^ {\dag})
Tr ( U +  U^ {\dag})
\nonumber \\
&+&
L_5 m_\pi ^2 Tr (\partial ^\mu U \partial _\mu U ^ {\dag} ( U +
U ^{\dag}  )) + L_6 m_\pi ^4 [Tr ( U +  U ^ {\dag})]^2
\nonumber \\
&+&
L_7 m_\pi ^4 (Tr ( U -  U ^ {\dag}))^2 + L_8 m_\pi ^4
Tr ( U^2 +  U^{\dag 2})
\end{eqnarray}
$L_i$ are low energy coupling constants which in principle can be determined
from QCD.
These coupling constants are obtained by comparison
with experiments such as $\pi \pi $ scattering .
Recently Fearing \cite{FS} has obtained an expression for ${\cal L}^{(6)}$
which  has some hundred terms.

Many years ago before a
systematic study of chiral perturbation theory as an effective theory for
low energy QCD was embarked upon, Skyrme \cite{Sk} proposed the following
Lagrangian as an effective theory of hadrons,

\begin{equation}
{\cal L}_{Sk} ={-F_0 ^2 \over 4} Tr(L_\mu L^\mu ) + {1 \over 32 e^2}
Tr[L_\mu\,,\,L_\nu]^2
\end{equation}
The first term in (26) is the well known non-linear $\sigma$-model
term eq.(21), and
the second term which is called Skyrme term is responsible for the soliton
solutions of the theory.

But, if one expands the Skyrme Lagrangian (26) in terms of pion fields it can
be seen that such interactions as $\pi ^0 \pi ^0 \to \pi ^0 \pi ^0 $
are forbidden \cite{PT1,PT2,B}. Now in the limit  $m_\pi \to 0 $, there are
only two possible independent quartic derivative terms. One of them is
the Skyrme term above,
and the other is $ {\gamma \over {8 e^2}} Tr (L_\mu L^\mu )^2 $ \cite{GL3}. By
adding this term to the Skyrme Lagrangian,
\begin{equation}
{\cal L}_{mod. Sk} ={-F_0 ^2 \over 4} Tr(L_\mu L^\mu ) + {1 \over 32 e^2}
Tr[L_\mu\,,\,L_\nu]^2 + {\gamma \over 8 e^2} Tr(L_\mu L^\mu)^2 \,,
\end{equation}
not only the
interaction $\pi ^0 \pi ^0 \to \pi ^0 \pi ^0 $ is now included but also the
accuracy of the quantitative results improve \cite{DG,PT1}.
Note that the coupling constants in the Lagrangian of the extended
Skyrme model (27) are related to the coupling constants $L_i$ in (25).

We are now in the position to develop
the chiral perturbation theory in the framework of non-commutative geometry.
As mentioned in section 2, the basic tools for model building in the framework
of non-commutative geometry are two things. First a suitable algebra $\cal A$
which describes
our geometrical space and second a k-cycle $({\cal H} ,D,\Gamma)$ which helps
us to develop a differential calculus on $\cal A $.

In choosing ${\cal A}$, we must be guided by the form of the one-form $L_\mu$
which appear in the ordinary ChPT,
\begin{equation}
L=L_\mu {\bf d}x^\mu = (U \partial _\mu U^{\dag }) {\bf d}x^\mu
\,\,\,\,\,\,\,\, , \,\,\,UU^{\dag }= U^{\dag } U = 1 \kern - .4em 1 .
\end{equation}
 and compare it with (8). It is then obvious that $ \gamma (L) $ in
non-commutative
version corresponds to $L$, for a single layer space,
\begin{equation}
\gamma (L) = \gamma ^\mu U \partial _\mu U ^{\dag } =U [ \partial \kern -0.5em
/ ,U^{\dag }]= U dU^{\dag} \,\,\,\,,U \in {\cal U}({\cal A})
\end{equation}
where $d$ is the exterior derivative on differential algebra. We should
therefore identify,
\begin{equation}
L = g dg^* \,\, ,
\end{equation}
with $g$ in a representation of ${\cal A}$.

Comparison with eq.(20) - (25) suggests that the simplest generalization of
the effective
lagrangian in non-commutative geometry up to
order $p^4$ should read:
\begin{eqnarray}
{\cal L}_{eff} & = &  K_1 Tr (g dg^* g dg^* )
+ K_2 Tr (g  dg^* g dg^* g  dg^* g dg^*)
\nonumber
\\
&+&  K_3 (Tr (g  dg^* g  dg^*))^2
\end{eqnarray}
where $K_i$  are the coupling constants of theory similar to $L_i$ (24) or
(25).
It is interesting to note that the first and
third order terms in powers of momenta,
vanish due to vanishing of the trace of odd number of Dirac matrices.

It is instructive to apply our non-commutative geometry machinery to
the
ordinary four dimensional manifold. For this purpose, we take ${\cal A}$
to be as in eq.(9), and consider the second order terms of the lagrangian (31)
only, then with $g=U(x) \in SU(N_f)$, we get,
\begin{eqnarray}
&{\cal L}_{eff}^{(2)}& =  K_1 Tr (g  dg^*\, g  dg^*)=
 (-K_1) Tr ( dg\, dg^*)
\nonumber \\
&=& (-K_1) Tr ([D,g] [D,g^*])
=  (-K_1) Tr(\gamma ^\mu \gamma ^\nu ) Tr (\partial _\mu U
\partial _\nu U^ {\dag})
\nonumber \\
&=& (-4 K_1) Tr(\partial _\mu U \partial ^\mu U^ {\dag})
\end{eqnarray}
This result is nothing but the 4-dimensional non-linear $\sigma $-model as we
expected. By comparison of  the lagrangian in (32) and (21),
\begin{equation}
K_1=- {F_0 ^2 \over 16}.
\end{equation}

Again before applying our method to the more general lagrangian of eq.(31)
let us confine ourselves to the first term still, but use the two layer
geometry of eqs.(10) and (11), with N the number of flavours and

\begin{equation}
g = \left( \matrix {U & 0 \cr 0 & U^{\prime} } \right) \,\,\,,
\,\,\,\,\,\, U,U^{\prime } \in U(N)
\end{equation}

Repeating the calculation
in (32) we get,
\begin{eqnarray}
&{\cal L}_{eff}^{(2)}& = (-4 K_1) Tr[ \partial _\mu U \partial ^ \mu
U ^{\dag} +(M U^\prime - U M ) (M^ {\dag} U^{\dag} - U^{\prime{\dag } }
M^{\dag} ) \nonumber \\
&+& \partial _\mu U^{\prime {\dag}} \partial ^\mu U^{\prime {\dag}}
+ (M^{\dag} U - U ^{\prime} M^{\dag}) (M U^{\prime {\dag}} - U^{\dag} M)].
\end{eqnarray}
For $M=0$ we simply get two independent non-linear $\sigma$-models
on two separate layers. In general, if we assume only

\begin{equation}
M M^{\dag} = M^{\dag} M =m^2 1 \kern - .4em 1,
\end{equation}
and set $U^{\prime } = 1 \kern - .4em 1 $, for simplicity, we get

\begin{equation}
{\cal L}_{eff}^{(2)}= (-4 K_1) Tr(\partial _\mu U \partial ^\mu
 U ^{\dag })+ (-2 K_1) m^2 Tr(U+U^{\dag } -2) ,
\end{equation}
which is the lagrangian of eq.(23), with
${F_0 ^2 m_\pi ^2 \over 4} Tr (U+U^{\dag }-2)$  the
symmetry breaking term. Thus we have generated the pion mass naturally.
According to Connes's  interpretation \cite{C2,C3}, the distance
between the layers of space
is proportional to $1 \over m$. So if we let the distance of these two layers
tend to infinity, the non linear $\sigma $-model is reproduced.

We will now take up the lagrangian to order 4 on the two layer space above,
\begin{eqnarray}
{\cal L}_{eff}^{(4)}&=&  K_2 Tr(g dg^* g dg^* g
 dg^* g  dg^*)+K_3 (Tr(g dg^* g dg^*))^2  \nonumber \\
&=& K_2 Tr ( dg  dg^* dg dg^*)) + K_3 (Tr (dg dg^*))^2 .
\end{eqnarray}
assuming  $U^{\prime }=1 $ again and eq.(36), a lengthy calculation leads to,
\begin{eqnarray}
{\cal L}_{eff} &=& -(4 K_1 + 32 K_2 m^2 - 96 K_3 m^2) Tr (\partial _\mu U
\partial ^\mu U^{\dag} )
\nonumber \\
&-&(2 K_1 + (64 K_2 + 16 K_3) m^4) Tr(U + U^{\dag})
+ (- 2 K_2 + 16 K_3) (Tr (\partial _\mu U \partial ^\mu U^{\dag })^2
\nonumber \\
&-&4 K_2 Tr (\partial ^\mu U \partial ^\nu U ^{\dag})  Tr (\partial _\mu U
\partial _\nu U ^{\dag})
+ 16 K_2 m^2 Tr(\partial ^\mu U \partial _\mu U^{\dag} \partial ^\nu \partial
_\nu U^{\dag })\nonumber \\
&+& 16 K_3 m^2 Tr(\partial _\mu U \partial ^\mu U^{\dag })
Tr(U+U^{\dag })
+ 16 K_2 m^2 Tr(\partial ^\mu U \partial _\mu U^{\dag } (U+U^{\dag }))
\nonumber \\
&+&4 K_3 m^4 (Tr(U+U^{\dag }))^2
+ 8 K_2 m^4 Tr (U^2 + U^{\dag 2 }) + {\rm constant }
\end{eqnarray}

Aside from the quadratic terms already recovered at the level of $p^2$,
we have therefore obtained all the terms in the ordinary ChPT to order
$p^4$, except for the $L_7$ term.
By comparing (39) with (25) and (23) we
may write the following
relations between the parameters of (39)
and physical parameters
$L_i, F_0$ and $ m_{\pi }$,
\begin{eqnarray}
{- F_0 \over 4} &=& 4 K_1 + 32 K_2 m^2 + 96 K_3 m^2 \\
{- F_0 \over 4} m_\pi ^2 &=& 2 K_1 m^2 + 64 K_2 m^4 +16 K_3 m^4 \\
L_1 &=& -2 K_2 +16 K_3 \\
L_2 &=& - 4 K_2 \\
L_3 &=& 16 K_2  \\
L_4 &=& 16 K_3 \left ( \matrix {m^2 \over m_\pi ^2 } \right ) \\
L_5 &=& 16 K_2 \left ( \matrix {m^2 \over m_\pi ^2 } \right ) \\
L_6 &=& 4 K_3 \left ( \matrix {m^4 \over m_\pi ^4 } \right )  \\
L_8 &=& 8 K_2 \left ( \matrix {m^4 \over m_\pi ^4 } \right )
\end{eqnarray}
It is to be noted that we have reproduced a whole series of terms in
the ChPT lagrangian, just starting from the simple form (31).
At the $\rho $ meson mass, $M_\rho = 770$ Mev, and decay constant
$F_0 =154$ Mev, we take $L_3 = (-4.4 \pm 2.5) \times 10^{-3}$,
and $L_1 = (0.7 \pm 0.3) \times 10^{-3}$ as inputs from \cite{GL2,EG} and
obtain the corresponding values for the remaining parameters,
\begin{eqnarray}
L_2 &=& (1.1 \pm 0.6)\times 10^{-3} \;,\;[\; (1.3 \pm 0.7) \times 10^{-3} \;]
\nonumber \\
L_4 &=& (0.1 \pm 0.5 )\times 10^{-3} \;,\;[\; (-0.3 \pm 0.5) \times
10^{-3}\;]\nonumber \\
L_5 &=& (-3.1 \pm 2.4 ) \times 10^{-3}\;,\;[\;(1.4 \pm 0.5) \times 10^{-3}\; ]
\nonumber \\
L_6 &=& (0.02 \pm 0.08) \times 10^{-3}\;,\;[\;(-0.2 \pm 0.3) \times 10^{-3}\;]
\nonumber \\
L_7 &=& 0\;,\; [\;(0.4 \pm 0.15) \times 10^{-3}\;] \nonumber \\
L_8 &=& (-1.1 \pm 1.1 ) \times 10^{-3}\;,\; [\;(0.9 \pm 0.3) \times 10^{-3}\;]
\end{eqnarray}
For the purpose of comparison we have also denoted the experimental values in
the brackets, taken from ref. \cite{GL2,EG}.
The agreement for $L_2, L_4, L_6 $ are good, while $L_5, L_7,$ and $L_8$ are
in disagreement.

Had we limited ourselves to the first two terms of our non-commutative
lagrangian eq. (31), i.e. $K_3=0$, and set the mass scale $m=0$, we would have
gotten the Skyrme model lagrangian, together with the non Skyrme term of
eq.(27),
\begin{equation}
{\cal L}^{(4)} = - 2 K_2 Tr ([L_\mu , L_\nu ]^2) + 4 K_2 Tr(L_\mu L^\mu)^2,
\end{equation}
which gives,
\begin{eqnarray}
K_2 &=& - {1 \over {64 e^2 } } \\
\gamma &=& - {1 \over 2 }.
\end{eqnarray}
This value of $\gamma $ agrees within the data at 1 Gev  energy within the
experimental error.
Surprisingly, it also agrees with the non-Skyrme term found by Anderianov
\cite{AA1} from bosonization method in QCD within a minus
sign\footnote{We would like to thank Maxim Polyakov for
clarification of this point.} (see also \cite{MR}).

\section *{Acknowledgements}

The authors wish to thank Hessameddin Arfaei, Shahin
Rouhani and Ahmad Shafei Deh Abad for fruitful discussions.

\newpage

\section *{Appendix}
\setcounter{equation}{0}
\renewcommand{\theequation} {A.\arabic{equation}}

An important feature of the linear $\sigma $-model \cite{GL} is its built
in mechanism of spontaneous symmetry breaking, which may indicate use of the
formalism of non-commutative geometry; and we will embark upon in this
appendix.
Recently Guo, et al \cite{G}, have used another formalism of non-commutative
geometry, \cite{Sit}, and constructed the linear $\sigma$-model in
that framework.

As in section 3, let us take the algebra ${\cal A}$ and Dirac operator $D$
as (10) and (11) respectively, with $N = 2$; then the one-form $\rho $
is,
\begin{equation}
\rho =\sum _i a^i \,\,[\,D\,,\,b^i\,] \,\,\,\,\,\,\,\,\,, a^i\,,\,b^i \in {\cal
A}
\end{equation}
where $a^i$ and $b^i$ are represented by,
\begin{equation}
a^i \to \,\,{\rm diag } \,\,(a^i_1,a^i_2) \,\,\,\,\,\,,\,\,\,\,\,
b^i \to \,\,{\rm diag } \,\,(b^i_1,b^i_2)
\end{equation}
and $a^i_j$ and $b^i_j$ are $2 \times 2 $ matrices.
Then by a straightforward calculation it is seen that,
\begin{eqnarray}
\rho =\left ( \matrix {A_1 & \gamma _5 \otimes \phi _{12} \cr \gamma _5 \otimes
\phi _{21} & A_2 }\right )
\end{eqnarray}
where
\begin{eqnarray}
A_m & = & \sum _i a_m ^i \,\,  \partial \kern -0.5em /  b_m ^i ,
\,\,\,\,\,m=1,2, \nonumber \\
\phi _{mn} & = & \sum _i a_m ^i (M b_n^i -b_m ^i M), \,\,\,\,\,m \neq n,
\end{eqnarray}
By condition $\rho = \rho ^*$,  $A_m$ and $\phi _{mn} $, satisfy
\begin{equation}
A_m^* =A_m \,\,\,\,\,{\rm and} \,\,\,\,\, \phi _{mn}^*= \phi _{mn}
\end{equation}
To find the curvature $\theta $, let us first calculate
\begin{equation}
d \rho = \sum_i da^i \,\, db^i =\sum _i \,\,[\,D\,,\,a^i\,]\,\,[\,D\,,\,b^i\,].
\end{equation}
Matrix elements of $d\rho $ are:
\begin{eqnarray}
(d\rho )_{11}= \partial \kern -0.5em /  A_1 +(M \phi _{21} + \phi_{12} M^{\dag
}) - X_1 \nonumber \\
(d\rho )_{12}=-\gamma _5 (\gamma ^\mu \partial _\mu \phi _{12} + A_1 M-M A_2)
\nonumber \\
(d\rho )_{21}=-\gamma _5 (\gamma ^\mu \partial _\mu \phi _{21} + A_1 M-M A_2)
\nonumber \\
(d\rho )_{22}= \partial \kern -0.5em /  A_2 +(M^{\dag } \phi _{21} + \phi_{12}
M) - X_2 \end{eqnarray}
where $X_1$ and $X_2$ are the auxiliary fields
\begin{equation}
X_m = \sum _i \gamma ^\mu \gamma ^\nu a_m ^i \partial _\mu \partial _\nu b_m ^i
+ a_m ^i [M M^{\dag } ,b_m ^i ] ,\,\,\,\,\,\, m=1,2,
\end{equation}
Using the standard methods \cite{Ch1} we eliminate these auxiliary fields and
obtain
the form of $\theta $ in terms of the gauge fields. We then use the Dirac
operator,

\begin{equation}
D=\left( \matrix{  \partial \kern -0.5em /  \,\, \otimes \,\, { 1\kern - .4em
1} \otimes \,\,{ 1\kern - .4em 1} & \gamma
_5 \otimes M \otimes K \cr \gamma _5 \otimes M^{\dag} \otimes K^{\dag} &
\partial \kern -0.5em /
 \,\, \otimes \,\, { 1\kern - .4em 1} \otimes \,\,{ 1\kern - .4em 1} }
\right) \end{equation}
and set $A_{i \mu }=0$ to restrict ourselves to the Higgs field

\begin{equation}
\theta = \left( \matrix{ { 1\kern - .4em 1} \otimes ( H H^{\dag} + m^2 )
\otimes K K^{\dag} &
- \gamma ^5 \gamma ^{\mu} \otimes \partial _{\mu} H \otimes K \cr
- \gamma ^5 \gamma ^{\mu} \otimes \partial _{\mu} H^{\dag} \otimes K^{\dag}&
 { 1\kern - .4em 1} \otimes ( H^{\dag} H + m^2 ) \otimes  K^{\dag} K } \right)
\end{equation}
where
\begin{equation}
H= \phi + M \,\,\in M_2(I \kern -.6em C) .
\end{equation}
One may expand $H$ in terms of
the Pauli matrices $\tau $
\begin{equation}
H=\sigma ^\prime 1 \kern - .4em 1 + i \pi ^\prime . \tau
\end{equation}
where $\pi ^\prime$ is a three component $(\pi ^\prime_1,\pi ^\prime_2,
\pi ^\prime_3)$ object.
Then for Yang-Mills action we obtain
\begin{equation}
S_{YM} = \int dx^4 \{-8 Tr(K K^{\dag}) [ ( \partial _ {\mu} \sigma ^\prime )^2
+ ( \partial _ {\mu} \pi ^\prime)^2 ] + 8 Tr(K K^{\dag})^2 (\sigma ^{\prime 2}
+ \pi ^{\prime 2 }+ m^2 )^2\}
\end{equation}
For the fermionic part of theory first we construct the operator $D+\rho $,
\begin{equation}
D+ \rho  = \left( \matrix {  \partial \kern -0.5em /  \otimes { 1\kern - .4em
1} \otimes { 1\kern - .4em 1} &
\gamma ^5 \otimes H \otimes K \cr \gamma ^5 \otimes H^{\dag} \otimes K^{\dag}
&  \partial \kern -0.5em / \otimes { 1\kern - .4em 1} \otimes { 1\kern - .4em
1} } \right)
\end{equation}
then taking the fermionic field $\Psi $ as a two component field,
\begin{equation}
\Psi =\left ( \matrix {\Psi _L \cr \Psi _R} \right )
\end{equation}
where $\Psi _L $ and $\Psi _R $ are the left and right handed spinor fields,
one can show that,
\begin{equation}
S_F = \int d^4 x \{ \bar \Psi _L \partial \kern -0.5em /  \Psi _L +
\bar \Psi _R  \partial \kern -0.5em /  \Psi _R + \bar \Psi _L (\sigma ^\prime +
i \tau . \pi ^\prime ) \Psi _R K +
 \bar \Psi _R (\sigma ^\prime + i \tau . \pi ^\prime ) \Psi _L K^{\dag} \}.
\end{equation}
Finally adding $S_{YM} $ and $S_F$, the total action is
\begin{eqnarray}
S = \int d^4 x \{ \bar \Psi _L \partial \kern -0.5em / \Psi _L +
\bar \Psi _R  \partial \kern -0.5em /  \Psi _R + \bar \Psi _L (\sigma  + i \tau
. \pi ) \Psi _R {K \over \alpha}  +
 \bar \Psi _R (\sigma  + i \tau . \pi ) \Psi _L  {K^{\dag}
\over \alpha } \nonumber \\ +1/2 [ ( \partial _\mu \sigma )^2 +
( \partial _ {\mu} \pi  )^2 ] -{\beta ^2 \over {4 \alpha ^2}} (\sigma ^2
+ \pi ^2 + m^2 )^2 \}
\end{eqnarray}
where
\begin{eqnarray}
-8\,Tr(K K^{\dag}) = {1 \over 2} \alpha ^2 \;\;,\;\;
8\, Tr(K K^{\dag})^2 = -{1 \over 4 } \beta ^2  \;\;,\;\;
\sigma  = \alpha \sigma ^{\prime}\;\;,\;\;
\pi  = \alpha \pi ^{\prime}
\end{eqnarray}
If we compare the above action with the standard action of the linear sigma
model \cite{PO}
\begin{eqnarray}
S_{L \sigma} = \int d^4 x \{ \bar \Psi _L \partial \kern -0.5em /  \Psi _L +
\bar \Psi _R \partial \kern -0.5em /  \Psi _R + g\,  \bar \Psi _L (\sigma  + i
\tau . \pi  ) \Psi _R  +
g\,  \bar \Psi _R (\sigma  + i \tau . \pi  ) \Psi _L
 \nonumber \\
+1/2 [ ( \partial _ {\mu} \sigma  )^2 +
( \partial _ {\mu} \pi  )^2 ] -{\lambda \over 4} (\sigma
^2 + \pi ^2 + {\mu ^{2} \over \lambda} )^2 \}
\end{eqnarray}
the parameters in the lagrangian (A.17) are
\begin{eqnarray}
g= {K \over \alpha} \;\;,\;\;
\lambda = {\beta ^{2} \over \alpha ^ {2} } \;\;,\;\;
{\mu ^{2} \over \lambda}=  \alpha ^2 m^2.
\end{eqnarray}
For pion decay constant and the nucleon mass we obtain:
\begin{eqnarray}
F_0 = \mid  \alpha ^2 m ^2 \mid ^{1 \over 2} \;\;{\rm and}\;\;
M_N = {F_0 \over 4}
\end{eqnarray}
reminiscent of the Goldberger-Treiman relation.

\end{document}